\documentclass{aa}  
\usepackage{graphicx}
\usepackage{physics,amsmath}
\usepackage{txfonts}

\usepackage{natbib}
\bibpunct{(}{)}{;}{a}{}{,} 
\usepackage[colorlinks=true, allcolors = blue]{hyperref}

\hypersetup{
  colorlinks   = true,
  urlcolor     = blue,
  linkcolor    = blue,
  citecolor   = blue 
}
\usepackage{xcolor}

\begin{document} 

   \title{An on-the-fly line-driven wind iterative mass-loss estimator (\href{https://lime.ster.kuleuven.be/}{LIME}) for hot, massive stars of arbitrary chemical compositions}
   \author{J.O. Sundqvist, D. Debnath, F. Backs, O. Verhamme, N. Moens, L. Delbroek, D. Dickson, P. Schillemans, C. Van der Sijpt, and M. Dirickx}

   \institute{Instituut voor Sterrenkunde, KU Leuven, Celestijnenlaan 200D, 
              3001 Leuven, Belgium,\\
    \email{jon.sundqvist@kuleuven.be}}

   \date{Received 2025-04-01; Accepted 2025-10-02}

   \titlerunning{mass loss calculator}
   \authorrunning{Sundqvist et al.}

  \abstract
   {Mass-loss rates, $\dot{M}$, from hot, massive stars are important for a range of astrophysical applications.} 
   {We present \href{https://lime.ster.kuleuven.be/}{LIME}, a fast, efficient, and easy-to-use real-time mass-loss calculator for line-driven winds from hot, massive stars with given stellar parameters and arbitrary chemical compositions. The tool is publicly available online.} 
   {We compute the line force on-the-fly from excitation and ionization balance calculations using a large atomic data base containing more than four million spectral lines. We then derive mass-loss rates from line-driven wind theory, including effects of a finite stellar disk and gas sound speed.} 
   {For a given set of stellar parameters and chemical composition, we obtain predictions for $\dot{M}$ and for the three line-force parameters, $\bar{Q}$, $Q_0$, and $\alpha$, at the wind critical point. A comparison of our predicted $\dot{M}$ with a large sample of recent, state-of-the-art, homogeneously derived empirical mass-loss rates obtained from the {\sc XshootU} collaboration project demonstrates that the simple calculator presented here performs on average as well as, or even better than, other available mass-loss recipes based on fits to restricted model grids computed from more sophisticated but less flexible methods.} 
   {In addition to its speed and simplicity, a strength of our mass-loss calculator is that it avoids uncertainties related to applying fit formulae to underlying model grids calculated for more restricted parameter ranges. In particular, individual chemical abundances can be easily modified, and their effects on predicted mass-loss rates can be readily explored. This enables direct applications also to stars that are significantly chemically modified at the surface. }

   \keywords{Stars: mass-loss - Stars: winds, outflows - Stars: massive - Methods: numerical - Methods: analytical}

   \maketitle
   
\section{Introduction}

For hot, massive stars, absorption and scattering in spectral lines transfer significant momentum from the star's intense radiation field to the plasma, providing a force that can overcome gravity and drive a strong, fast wind outflow. Line-driven wind theory based on this mechanism was established in the seminal paper by \citet{cak1975} (hereafter CAK). Earlier attempts \citep[e.g.,][]{Lucy70} focused on obtaining the line force from a small number of select lines, whereas CAK used results obtained for carbon to parametrize the cumulative line acceleration stemming from numerous lines. While various extensions and reformulations of this elegant model have substantially improved its quantitative applicability as well as our physical understanding \citep[e.g.,][]{Abbott82, castor_friend_1983, owocki_1984, pauldrach_1986, owocki88, Kudritzki89, gayley_1995, Cranmer95, puls_2000, Kudritzki00, Kudritzki02, cure04, Lattimer21, luka_2022, Key25}, many key elements of line-driven wind theory were undoubtedly already established by CAK. 

Much effort has been devoted to obtaining quantitative predictions for mass-loss rates, $\dot{M}$, suitable for direct comparisons with observations and for applications such as stellar evolution and feedback. Typically, models in this category have used detailed radiative transfer techniques applied directly to large discrete line lists; examples include models based on the co-moving frame \citep[CMF;][]{pauldrach_1986, sander_2017, krticka_2017, sundqvist_bjorklund_2019} 
as well as Monte Carlo \citep{Vink00, lucy10} radiative transfer. Since these calculations are very time-consuming, a commonly used approach has been to compute relatively small grids of models and then design ``mass-loss recipes'' by fitting grid results to some assumed functional behavior of $\dot{M}$ with stellar parameters \citep[e.g.,][]{vink_2001, Sander20, robin_2021, Bjorklun23, Krticka24}. 
Although these approaches have been very useful, the resulting recipes are very inflexible because the underlying model-grid calculations are computationally intensive. 
Consequently, many applications in practice rely on (sometimes questionable) extrapolations to domains outside the actual model grids. Moreover, differences between actual model-grid results and the applied fit formulas introduce significant scatter and additional uncertainty. Another important aspect of the effort here is that, in general, it has not yet been possible to directly examine the effects of varying individual chemical abundances on line-driven mass-loss predictions, although a few targeted attempts have been made (e.g., \citealt{Vink05, Krticka19}). In practice, this means that applications often rely only on simple scaling relations with overall metallicity \citep[e.g.,][]{Meynet06}.  

Building on line-driven wind theory, we present a fast and reliable method for directly computing line-driven mass-loss rates from stars with given stellar parameters and arbitrary abundance patterns. 
We tested this method by comparison with state-of-the-art observational results from the large {\sc XShoot Ullyses}
project \citep{Vink23}, showing that our method performs better than several more elaborate but less flexible mass-loss recipes that are currently available. Motivated by these results, we provide a new, user-friendly online mass-loss calculator \href{https://lime.ster.kuleuven.be/}{LIME}, which accepts arbitrary stellar parameters and chemical compositions (within our recommended limits, see Sect. \ref{discussion}). On-the-fly calculations typically take only a few seconds, and users can also submit larger batch jobs. 
Our method is therefore well suited for direct applications to observed stars with specific parameters, as well as for broader applications such as stellar evolution and feedback. 

\section{Method and applicability}

\begin{figure}
 \centering
 \includegraphics[width=9cm]{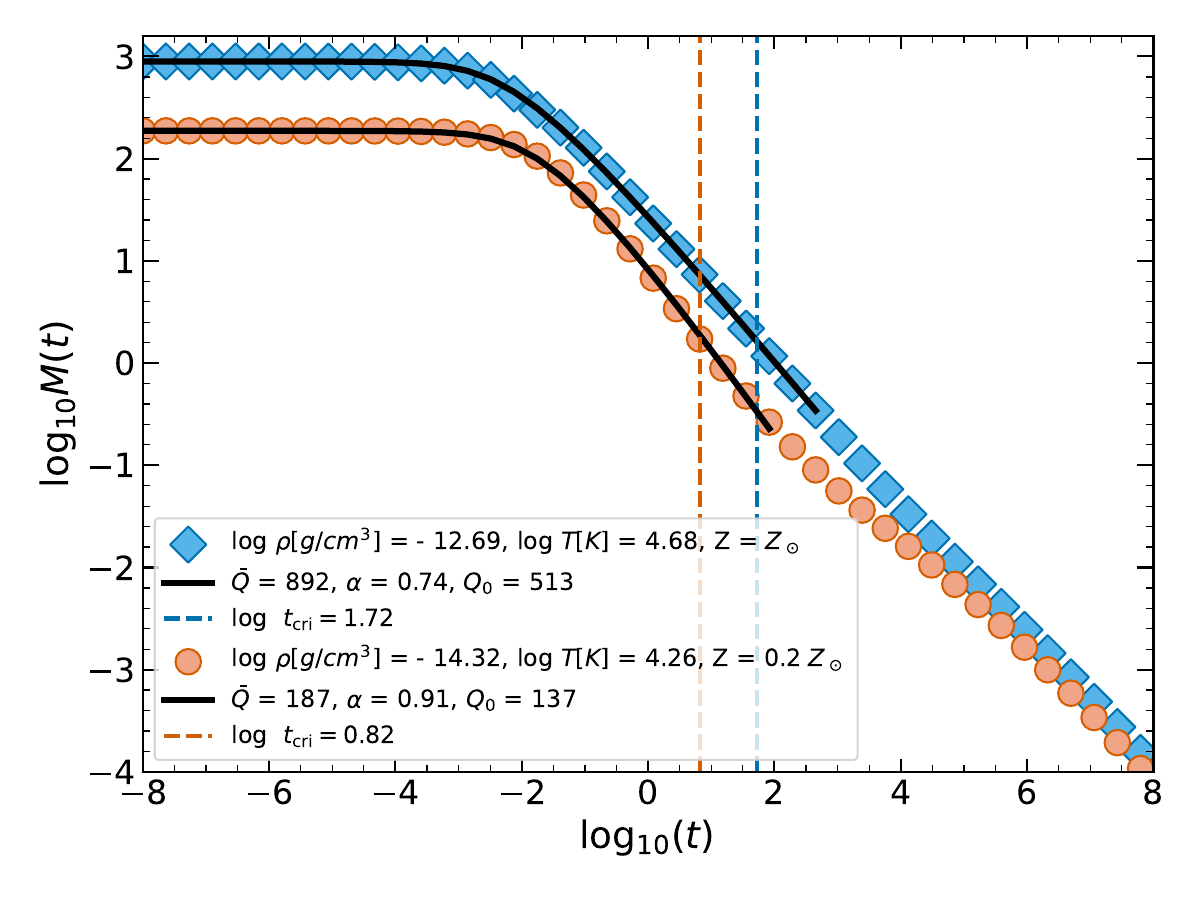}
      \caption{Line-force multiplier, $M(t)$, as a function of the Sobolev optical-depth-like parameter, $t$. The blue diamonds and orange circles show the line-force multiplier calculated for the characteristic density, temperature, and metallicity listed in the labels. Black lines indicate the best fits with the corresponding line-force parameters. Relative metal abundances are scaled to the solar metallicity of \citep{asplund_2009}.}
  \label{Fig:Mfit}
\end{figure}

To compute mass-loss rates we applied CAK theory as formulated by \citet{gayley_1995}, modified for a high-opacity cut-off in line strength \citep{owocki88}, including the finite-disk angle effect \citep{pauldrach_1986} according to \citet{Kudritzki89}, and a finite sound speed correction following \citet{Owocki04b}. The analytic mass-loss rate as derived from the modified CAK critical point analysis is therefore 
\begin{equation} 
	\dot{M} = \frac{L}{c^2} \frac{\alpha}{1-\alpha} \left( \frac{\bar{Q} \Gamma_e}{1-\Gamma_e} 
	\right)^{1/\alpha-1} \frac{\bar{Q}}{Q_0} \left( \frac{1}{1+\alpha} \right)^{1/\alpha} \left(1 + \frac{4 \sqrt{1-\alpha}}{\alpha} \frac{a}{\varv^{\rm eff}_{\rm esc}}\right), 
    \label{Eq:mdot}
\end{equation}  
as given, for example, by \citet{gayley_1995} (their Eq. 43), applying a finite-disk correction factor from \citet[][Eq. 32]{Kudritzki89} and a sound-speed correction as outlined in Appendix A of \citet{Owocki04b}. Here, $L$ is the stellar luminosity, $c$ the speed of light, $\Gamma_{\rm e} \equiv g_{\rm e}/g = L \kappa_{\rm e} / (4 \pi c G M)$ the classical Eddington ratio for electron scattering opacity $\kappa_{\rm e}$ and stellar mass $M$, $a$ is the isothermal gas sound speed, and $\varv^{\rm eff}_{\rm esc}$ the escape speed from the stellar surface, reduced by the contribution from $\Gamma_{\rm e}$. The line-force parameters are $\bar{Q}$, $Q_0$, and $\alpha$. For a given set of stellar parameters, the present problem is thus reduced to finding appropriate values for these and for $\kappa_{\rm e}$ at the wind critical point. 

The line acceleration, $g^{\rm tot}_{\rm line}$, is computed from the sum 
\begin{equation} 
	\frac{g_{line}^{tot}}{g_e} \equiv M(\rho,T,t) = \sum_{i}^{N} w_{\nu,i} q_i 
	\left( \frac{1- e^{-q_i t}}{q_i t} \right),   
	\label{Eq:msum} 
\end{equation}
for a line list consisting of $N$ lines, each indexed by $i$, so that $M(t)$  defines the line-force multiplier with respect to the electron-scattering acceleration, $g_e$. 
Here $q_i = k_{L,i} / (\nu_0 \rho \kappa_{\rm e})$ is the normalized line strength for the frequency-integrated line extinction coefficient $k_L \, \rm [cm^{-1} Hz]$ and line-center frequency $\nu_0$, $w_{\nu,i}$ is the frequency-dependent flux weighting with fluxes approximated as Planckian distributions \citet{luka_2022}  (see also \citealt{puls_2000}), and $q_i t = \tau_i$ is the line optical depth $\tau_i$ \citet{sobolev_1960}. Importantly, since $t = \kappa_e c \rho / \left|d \varv /dr\right|$ is independent of the line index $i$, it can be treated as an independent variable of the problem \citep{cak1975}. For further discussions and full derivations of these quantities, we refer to \citet{luka_2022}, in particular Sects. 2.1-2.2. As in \citet{luka_2022}, we computed this sum for a range of $t$ by solving the excitation and ionization balance for a given mass density, $\rho$, and temperature, $T$, using the Munich atomic line database \citep{pauldrach_2001, puls_2000, puls_2005}\footnote{This database consists of $>$ 4 million lines covering all elements from hydrogen to zinc (except Li, Be, B, and Sc, which are too rare to affect the accumulative line-opacity) and ionization stages up to VIII. It is the same database as included in the model-atmosphere code {\sc fastwind}
(see \citealt{puls_2005}, their Sect. 3).} and following the Saha-Boltzmann procedure outlined in \citet{luka_2022} (see also \citealt{Lattimer21}). We then fit the result to
\begin{equation} 
	M_{\rm fit}(\rho,T,t) = \frac{\bar{Q}}{(1-\alpha)}
	 \frac{(1+Q_0 t)^{1-\alpha}-1}
	 {Q_0 t}. 
	 \label{Eq:mfit} 
\end{equation}
This functional form follows from the line-distribution function of \citet{gayley_1995} (see Sect. 2.3 in \citealt{luka_2022}). For a given $(\rho,T)$ pair, the fit provides values of $Q_0$ and $\alpha$, while $\bar{Q}$ can be computed directly from the sum in Eq. \ref{Eq:msum}; representative examples are shown in Fig. \ref{Fig:Mfit}. The quantity $\bar{Q} \equiv \sum_i w_{\nu,i} q_i$ sets the saturation for the force multiplier in the limit that all lines are optically thin. The parameter $Q_0$ is related to the position where the optically thick lines begin to appear, and $\alpha$ determines the slope of $M(t)$ in the region where these effects are visible\footnote{In their study, CAK considered only this power-law component.}. The physical meaning of these parameters is discussed in \citet{gayley_1995, puls_2000, luka_2022}.\footnote{Because we obtain a new excitation and ionisation balance for each pair of $(\rho,T)$, there is no need to introduce additional parameters, such as \citealt{Abbott82}'s $\delta$-parameter (see \citealt{luka_2022}) when fitting $M(t)$ using Eq. \ref{Eq:mfit}.}  
\begin{figure}
 \centering
 \includegraphics[width=9cm]{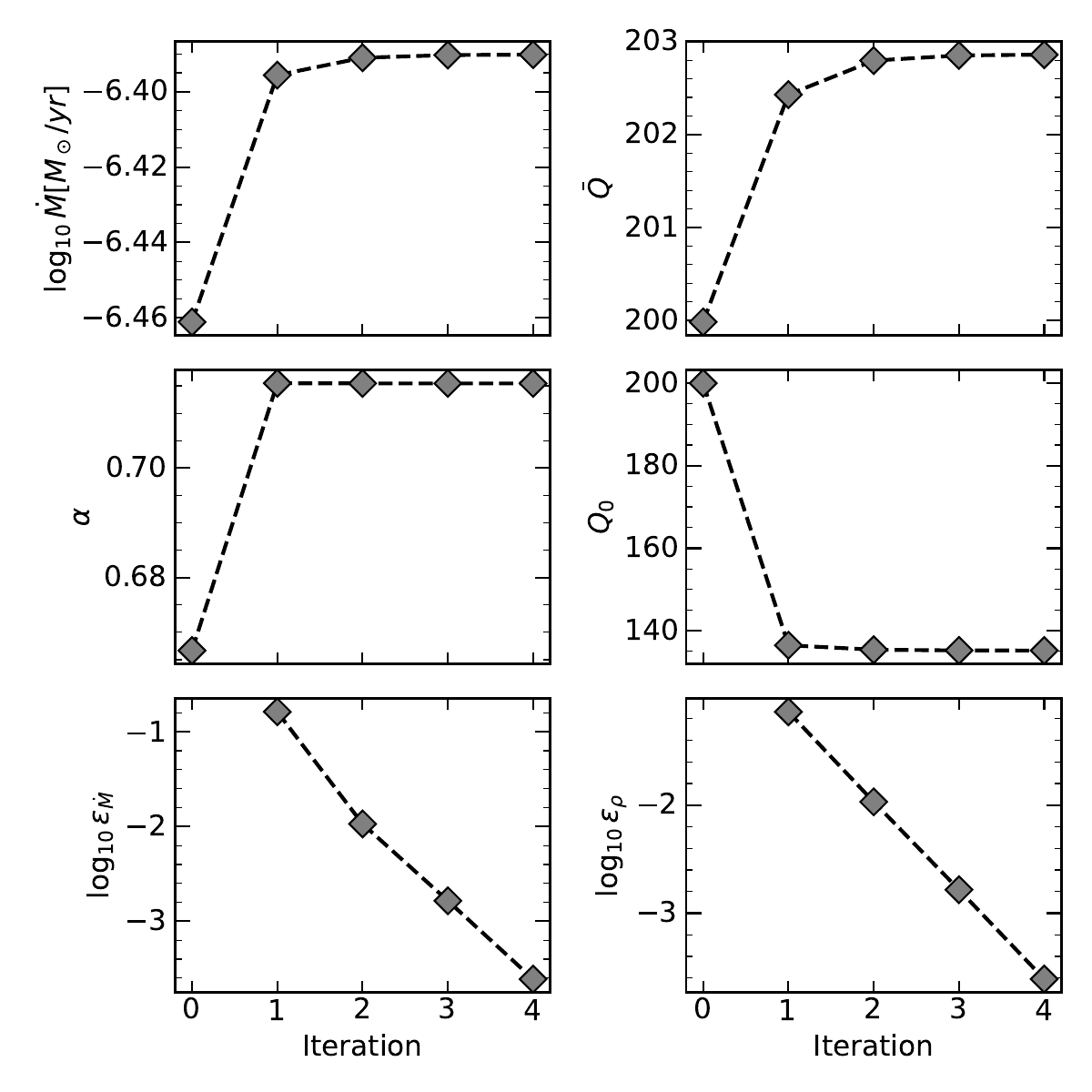}
      \caption{Quantities $\dot{M}$ (top left), $\Bar{Q}$ (top  right), $\alpha$ (middle  left), and $Q_0$ (middle  right) as a function of iteration number until convergence (after 3 iterations). The bottom panels show relative differences in $\dot{M}$ (left) and critical point density $\rho$ (right) between successive iterations.}
  \label{Fig:iteration}
\end{figure}
We evaluate the line-force parameters at the wind critical point and inserted into the mass-loss rate formula above. Ionization and recombination keep the wind approximately at the  photospheric temperature \citep{puls_2000} so we set $T = T_{\rm eff}$ for our calculations (see also \citealt{Lattimer21}). The critical point density, $\rho_c$, is initially unknown. For a critical point velocity, $\varv_c$, and radius, $r_c$, we compute $\rho_c = \dot{M}/(4 \pi \varv_c r_c^2)$. We apply the finite-disk effect \citep{pauldrach_1986}, which places the wind critical point very close to the stellar 
surface, $R_\ast$, so that $r_c = R_\ast$ is generally a good approximation. For the critical velocity $\varv_c$ we apply the modified CAK ``cooking recipe'' of \citet{Kudritzki89} (see their eqns. 55, 55a). 

For each set of stellar parameters and chemical composition, we first guess values for the line-force parameters, applying Eq. \ref{Eq:mdot} and computing an initial estimate of $\rho_c$. 
Using this estimate, we solve for the sum in
Eq. \ref{Eq:msum} for a range of $t$, fit Eq. \ref{Eq:mfit} to the results and obtain updated values for the line-force parameters. We then use the updated parameters to compute a new mass-loss rate and repeated the procedure until $\rho_c$ (and hence $\dot{M}$) converged. This procedure yields a simple, robust, fast, and efficient scheme that typically achieves good precision after very few iteration steps (often only about four; see Fig. \ref{Fig:iteration}). The results are predictions for $\dot{M}$ and values of $\bar{Q}$, $Q_0$, and $\alpha$ at the wind critical point for each set of stellar input parameters. A typical example of the iteration is shown in Fig. \ref{Fig:iteration}. 
An interactive version of the procedure is available at 
\href{https://lime.ster.kuleuven.be/}{LIME}\footnote{https://lime.ster.kuleuven.be}.

\begin{figure}
 \centering
 \includegraphics[width=9cm]{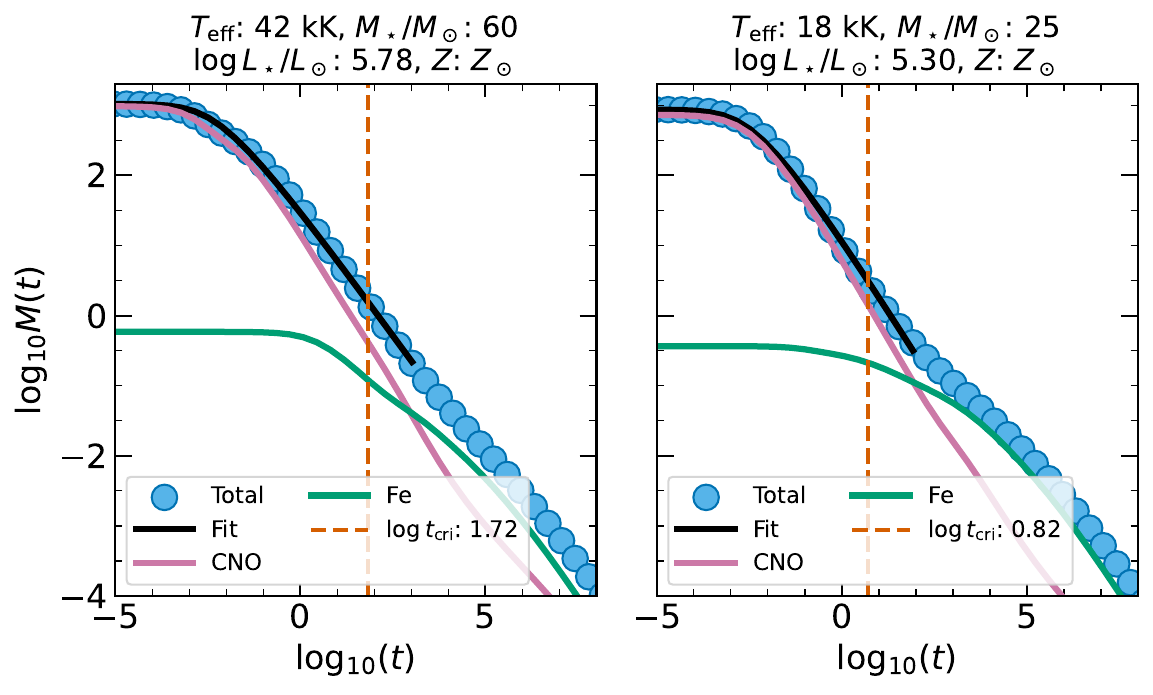}
      \caption{Line-force multiplier, $M(t)$, as a function of $t$ for stars with $T_{\rm eff}$ and critical point density (log $\rho_{\rm crit} \ [gcm^{-3}]$) of 42000 [K] and -12.79 (left) and 18000 [K] and -14.52 (right). The line-force parameters $\bar{Q}$, $Q_0$, and $\alpha$ at the critical point are 1043, 620, 0.73 (left) and 866, 834, 0.87 (right). Blue circles represent $M(t)$ calculated from the entire line list, with the black line indicating the fit. Green and pink lines show $M(t)$ calculated using only Fe lines and CNO lines, respectively. The vertical dashed orange line marks $t_{\rm cri}$ (see text).}
  \label{Fig:Mcompare}
\end{figure}

Occasionally, different parts of the $M(t)$ curve for a given $(\rho,T)$ pair have different slopes. This typically occurs at relatively low temperatures, as shown in Fig. \ref{Fig:Mfit} (orange circles). Fig. \ref{Fig:Mcompare} demonstrates that this behavior relates to the contributions of different driving ions to the total line force. In the example shown, the iron line contribution (green curve) displays a shallower slope than the carbon, nitrogen, and oxygen (CNO) lines and a significantly lower saturation limit, $\bar{Q}$. While iron contributes significantly to the total line force at higher $t$ values, CNO dominates $M(t)$ at lower $t$, which is then also reflected in the slope of the cumulative curve. This slope difference between light (CNO) and heavy (iron) elements was also found by \citet{puls_2000}. To ensure that we covered the ``correct'' part of the $M(t)$ curve, we estimated $t_{\rm cri} = \kappa_e \rho_{\rm cr} c /|dv/dr_{\rm cr}|$ using the analytically predicted velocity gradient for the critical point in the CAK radial stream limit. 
When fitting for $Q_0$ and $\alpha$, we then ensured that the fitting range included both $t_{\rm cri}$ and the saturation limit. The latter is particularly important because saturation occurs at different $t$ values in different parts of the parameter space (for example, depending on metallicity, $Z$).

A reasonable lower limit for line-driven winds is $\Gamma_e (1 + \bar{Q}) \ga 1$, since below this the radiation force is never able to overcome gravity\footnote{Gas pressure gradients are typically small in regimes where line-driving is important}. As such, we imposed $\Gamma_e (1 + \bar{Q}) =1$ as a lower limit for our models, but note that a somewhat stricter limit of the present theory could be $M(t_{\rm cri}) < \bar{Q}$. If this criterion is not met, the line-force multiplier does not depend on the velocity gradient, rendering the CAK-like critical point analysis leading to Eq. \ref{Eq:mdot} questionable. Although not relevant for the observational comparisons presented in Sect. \ref{comparison}, these lower limits can be important for cases with low $\Gamma_{\rm e}$ and in the low-metallicity regime (see also \citealt{Kudritzki02}), owing to the principal dependence $\bar{Q} \sim Z$. Finally, we note that we did not consider possible decoupling of metal ions from the bulk hydrogen and helium plasma, which may complicate the situation for very low-density winds (see \citet{Owocki02} for suitable estimates of where this effect might become relevant). 

The present version of \href{https://lime.ster.kuleuven.be/}{LIME} evaluates the line force parameters only at the estimated wind critical point, which determines the mass-loss rate within the formalism. As such, the effects of relative differences in line-driving efficiency between the inner and outer wind regions are neglected (see also Sect. \ref{discussion}). However, we can still use the line-force parameters calculated by \href{https://lime.ster.kuleuven.be/}{LIME} to obtain also an approximate value for the terminal wind speed $\varv_\infty$. For this, we adopted the fitting-formulas  of \citet[] [Eqs. 15 and 17, assuming $\delta =0$]{Kudritzki00}. As noted above, because our method does not take into account differences in line-driving efficiency between the wind critical point and the outer wind, terminal wind speed estimates are generally expected to be less precise than our corresponding $\dot{M}$ calculations (see, e.g., \citealt{puls_2000, robin_2021} for discussions of how outer-wind effects may influence $\varv_\infty$). When comparing our theoretical $\varv_\infty$ estimates with values from the empirical studies described below, we find reasonably good agreement for early supergiants, whereas theoretical values for dwarfs are sometimes higher than empirically inferred $\varv_\infty$-- a general result that has been noted several times before in the line-driven wind literature (e.g.,\citealt{Kudritzki00, robin_2021}). 

\section{Comparison with observations and other theoretical mass-loss rates}
\label{comparison}

\begin{figure}
 \centering
 \includegraphics[width=9cm]{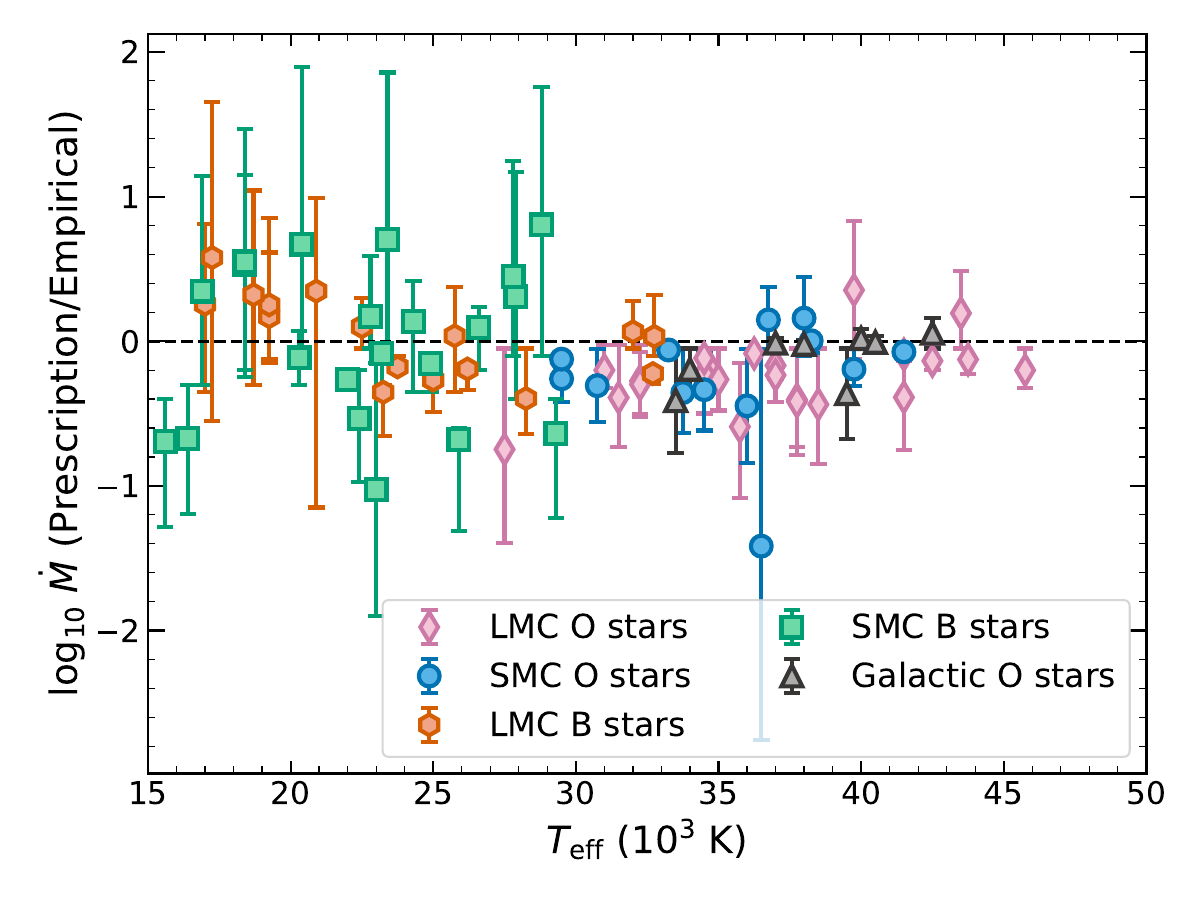}
      \caption{Logarithm of the ratio between mass-loss rates derived using our method and rates obtained empirically from observational data as described in the text. The colors and markers correspond to the different samples marked on the labels. The dashed line illustrates the position where rates would be equal.} 
  \label{Fig:cmp}
\end{figure}

\begin{figure}
 \centering
 \includegraphics[width=9cm]{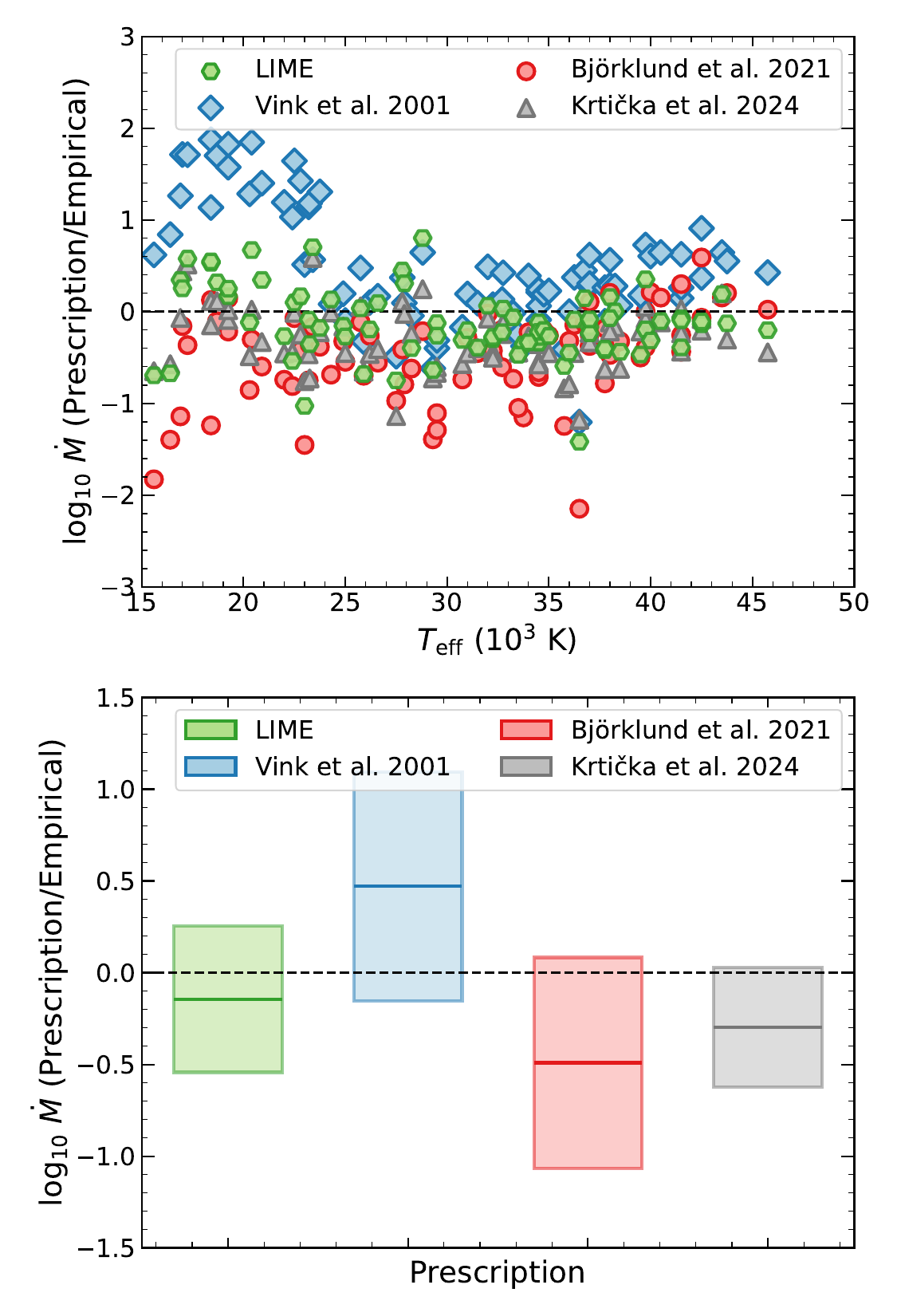}
      \caption{Top panel: Same as Fig. \ref{Fig:cmp}, but also comparing the observations with theoretical rates computed by different methods, as indicated in the labels. For clarity, error margins for empirical rates are omitted. Bottom panel: Sample averages and one standard deviation for the four methods. The dashed black line indicates where rates are equal.}
  \label{Fig:vink_robin}
\end{figure}

Fig. \ref{Fig:cmp} compares our predicted mass-loss rates with those derived from observations of 86 OB stars in the Galaxy, the large Magellanic cloud (LMC), and small Magellanic cloud (SMC). These empirical rates were derived homogeneously by fitting optical and ultraviolet observations of a multitude of strategic spectral lines to synthetic spectra computed with the model atmosphere and wind code {\sc fastwind} \citep{puls_2005, sundqvist_2018b}. This modeling accounts for wind clumping with arbitrary optical thickness, a non-void interclump medium, and porosity effects in velocity space.  Observational data were taken from the {\sc ulysses} and {\sc XshootU} projects \citep{Vink23} for the LMC and SMC, and from \citet{hawcroft_2021} for the Galaxy. For each star, we used a genetic algorithm approach to find the global best fit to the data \citep{hawcroft_2021,brands_2022}. These state-of-the-art empirical results are described in detail by \citet{hawcroft_2021, Verhamme24, Backs24, Brands25}, and Verhamme et al. (in preparation).

Fig. \ref{Fig:cmp} shows overall good agreement between our predicted rates and those empirically inferred from observations. The error bars indicate 1$\sigma$ statistical uncertainties in the empirical $\dot{M}$.  Although the comparison exhibits significant scatter, there appears to be no significant systematics present. Comparing our predictions with alternative mass-loss rate recipes from \citet{vink_2001, Bjorklun23} and \citet{ Krticka24}, shown in Fig. \ref{Fig:vink_robin} reveals that the simple method presented here performs equally well or better. For example, the systematics seen in the comparisons with \citet{vink_2001}, which predicts significantly higher rates at lower temperatures, and with \citet{Bjorklun23}, which systematically yields lower rates, were not present in our results. The prescription of \citet{Krticka24} also shows no clear systematics, but we note that this recipe is based on a complex fitting function formula applied to an underlying model grid. This makes its application outside the limited parameter domain ($\log_{10} L/L_\odot = 5.28-5.88$ and $Z/Z_\odot = 0.2-1.0$) difficult and not recommended (see discussions in \citet{Krticka24} and \citet{Verhamme24}). Sample means and standard deviations of all four comparisons yield the ratios $\log_{10} \dot{M}_{\rm prescription}/\dot{M}_{\rm empirical} = -0.12 \pm 0.41, 0.47 \pm 0.64, -0.53 \pm 0.56$, and $-0.30 \pm 0.34$ for the present method and those of \citet{vink_2001}, \citet{Bjorklun23}, and \citet{Krticka24}, respectively; these sample means and standard deviations are also plotted in the lower panel of Fig. \ref{Fig:vink_robin}. 
It remains unclear whether the remaining small offset and scatter in the comparisons between our method and the empirical data are dominated by uncertainties in the line-driven wind theory presented here or in the methods used to infer $\dot{M}$ from observational data (see also the discussions in \citealt{robin_2021, Verhamme24}). 

\begin{figure*}
 \centering
 \includegraphics[width=18cm]{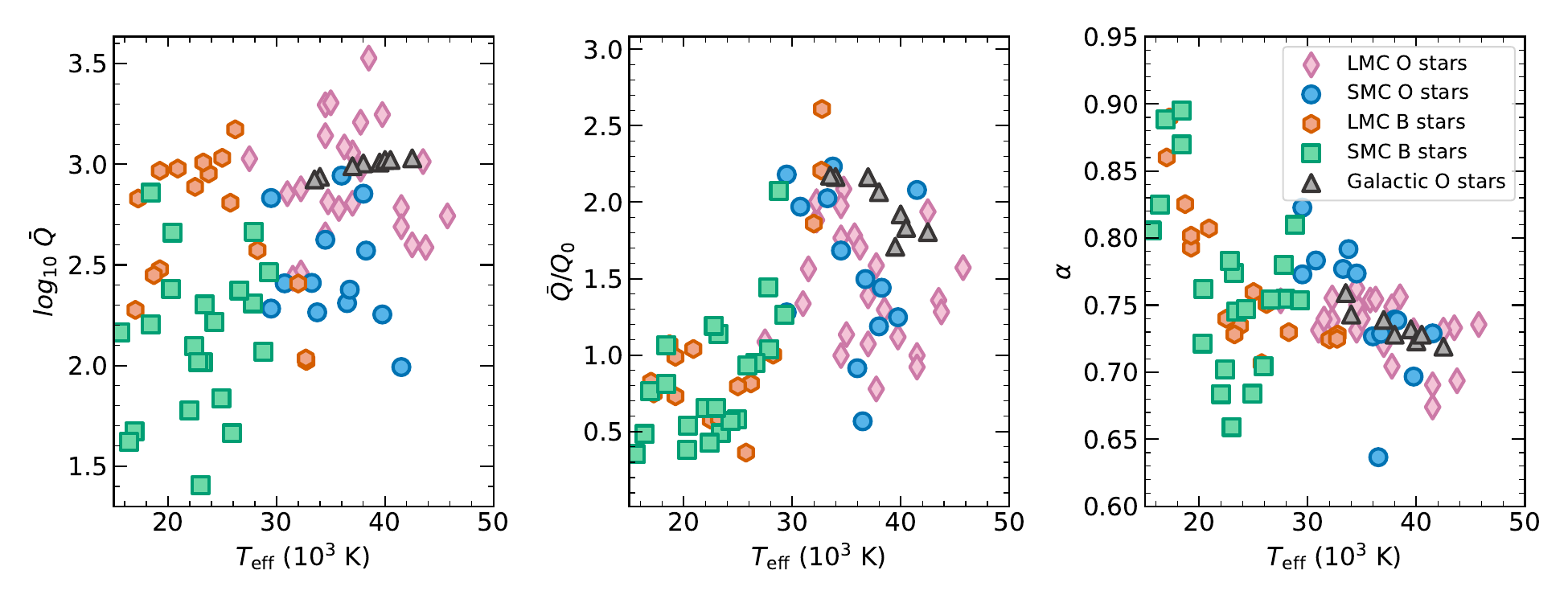}
      \caption{Line-force parameters $\Bar{Q}$ (left), $\Bar{Q}/Q_0$ (middle), and $\alpha$ (right) at the wind critical point for the stellar sample described in Sect. \ref{comparison}. Colors and markers correspond to the different stellar types indicated in the labels.}
  \label{Fig:lineforce_scatter}
\end{figure*}

We also provide the line-force parameters ($\alpha$, $\bar{Q}$, and $Q_0$) for the stars in our comparison sample (see Fig. \ref{Fig:lineforce_scatter}). As expected, the $\bar{Q}$ values are on average higher for LMC stars than for SMC stars because of the higher metallicity of the LMC; however, significant star-to-star scatter is present. We also note an interesting feature of large $\alpha$ values at lower temperatures.
The explanation for this is the same as that discussed in Fig. \ref{Fig:Mcompare}: whereas the slope of the $M(t)$ curve is shallower in the iron-dominated high-$t$ regions, $t_{\rm cri}$ for our sample stars typically falls within the steeper region influenced by lighter CNO ions, resulting in higher characteristic $\alpha$ values. This trend contrasts previous discussions in the line-driven wind literature regarding which ions primarily set $\dot{M}$. Specifically, while such discussions \citep[e.g.,][]{puls_2000} primarily focused on the role of CNO in outer wind driving at Galactic metallicities, we find that the steeper slopes of these lighter ions are often already reflected in the accumulative line force at $M(t_{\rm cri})$. 

\section{Discussion and conclusion}
\label{discussion}

The observational comparisons presented above demonstrate that our simple mass-loss calculator performs equally well or better than alternative recipes based on more elaborate line-driven wind models. This may partly arise because those recipes are fit formulas, typically assuming a power-law scaling of $\dot{M}$ with stellar parameters, based on underlying grid calculations for a restricted number of models. This introduces scatter and uncertainties when applied to actual stars (see Sect. \ref{comparison}). In contrast, our method computes all quantities on-the-fly for any given set of input stellar parameters, thereby avoiding such intrinsic uncertainties. Moreover, with our approach, we can directly tune individual abundances, whereas \citet{vink_2001, Bjorklun23, Krticka24} provide only general scalings with overall metallicity $Z$. As an example of this advantage, although the general scaling $\bar{Q} \sim Z$ holds \citep{gayley_1995}, stars with significant surface chemical modifications do not always follow this. Secondary effects from corresponding changes in $Q_0$ and $\alpha$ can also affect the final $\dot{M}$ scalings with metallicity differently, depending on the stellar parameter range (see also \citealt{puls_2000}). We anticipate this to be especially relevant for future applications to chemically modified stars in the early Universe \citep[e.g.,][]{Meynet06} and to stars stripped by binary interactions \citep[e.g.,][]{Gotberg23}. 

Overall, \href{https://lime.ster.kuleuven.be/}{LIME} yields better agreement with empirical studies than the theoretical rates we previously computed from the same atomic database. These rates were calculated using more sophisticated CMF radiative-transfer methods coupled with a full solution to the stationary equation of motion (\citealt{sundqvist_bjorklund_2019}, with corresponding grid calculations by \citealt{robin_2021, Bjorklun23}). In these CMF-based models, a local reduction in radiative acceleration around the gas sonic point, generally related to a source function dip (see \citealt{Owocki99}), effectively limits the mass flux that can be driven. This is in contrast to models using \citet{sobolev_1960} line transfer and results in overall lower predicted mass-loss rates, particularly for dwarf stars. Furthermore, inclusion of ad hoc so-called ``microturbulence'' in the broadening of line profiles significantly impacts the predicted line force in such non-Sobolev models (see \citealt{lucy10, robin_2021, Gemma25b}). Recent time-dependent multidimensional simulations of O stars reveal strong turbulence in the sonic regions \citep{Debnath24}. However, our results suggest that including this turbulence using the current standard ``microturbulence'' framework in stationary CMF-based models may not be justified, at least not when computing the line acceleration. In Appendix A, we further compare force multipliers computed with the Saha Boltzmann method used here to those obtained with the approximate NLTE method of \citet{puls_2000}. For a few select temperatures at a prototypical critical point density, we derived line-force parameters using both methods and find generally small differences. The main reason is that the mass-loss rate is determined very close to the stellar surface. At these radii, the densities are still relatively high and the effects of NLTE modest\footnote{This applies to the evaluation of cumulative line acceleration, but may not hold for detailed diagnostics of specific spectral lines (see Appendix A and \citealt{puls_2005})}. In the outer wind, NLTE effects are typically much larger due to the very diluted stellar radiation field. This underlies why \href{https://lime.ster.kuleuven.be/}{LIME} provides better results for $\dot{M}$ than for $\varv_\infty$ (see the end of Sect. 2). Specifically, our method
uses insights from line-driven wind theory to avoid explicit numerical solutions of the equation of motion, thereby obtaining mass-loss rates in a simplified way. It therefore computes the line force at a single (critical) point in the wind. This method can then provide on-the-fly quantitative predictions for mass-loss rates, since computing the sum in Eq. \ref{Eq:msum} 
is the only time-expensive step in the algorithm. In reality, however, density and temperature vary throughout the wind, so that the excitation and ionization balance, and thus the line-force parameters $\bar{Q}$, $Q_0$, and $\alpha$, also vary (see \citealt{luka_2022, nico_2022b, Debnath24} for examples of these variations within this formalism). This also applies to NLTE corrections, which generally increase with wind radius. These effects influence the analytic estimates of the wind-critical position and velocity applied here (see also \citealt{Kudritzki89, puls_2000}). However, the comparisons in Sect. \ref{comparison} and Appendix A suggest that they do not, in general, significantly affect the calculations of global $\dot{M}$ from line-driven winds.

Due to limitations in our atomic database, we currently do not recommend applying the recipe outside the range $18\,$kK\,$\la$$\,T_{\rm eff}\la 60\,$kK. We are currently working on an extension to lower and higher temperatures and will report on this in the near future (Debnath et al., in prep.). Such extensions are important, for example, for evolved hot stars whose surface layers have been stripped by binary interactions \citep{Gotberg23}.

In addition to these temperature restrictions,  a few further limitations of our method are worth noting. (i) Although the calculator can technically be applied up to the limit $\Gamma_e \rightarrow 1$, the theory is based on line-driving from the stellar surface, which means subsurface physical effects may complicate applications for stars residing very close to the classical Eddington limit, at least at (near) Galactic metallicities and above (see also
\citealt{Bestenlehner20}). Recent multidimensional simulations of evolved stars with $Z = Z_\odot$ show that above a certain $\Gamma_{\rm e}$ threshold, opacities from ionization of iron-group elements around $T \sim 150$ kK \citep{iglesias_1992} provide enough radiation force to launch a strong wind outflow from beneath the stellar surface, leading to enhanced mass-loss rates if the wind is also sustained to higher radii \citep{nico_2022a}. Similar results have also been found for recent 1D, stationary models \citep{sander_v_2020}. However, this effect is strongly metallicity dependent, so that hot stars with significantly lower abundances of iron-group elements can reside much closer to the Eddington limit with undergoing subsurface wind launch. (ii) We neglect the effects of rotation; for fast-rotating stars, corrections may be required (see Sect. 2.2.1 of \citealt{puls08}). (iii) We also neglect the influence of strong magnetic fields, which, at least for a small subset of massive stars, can induce magnetospheres that prevent large portions of the initiated wind from escaping the star (see \citealt{ud_doula08}, their Eq. 23, for a suitable scaling relation). 

\begin{acknowledgements}

The authors gratefully acknowledge support from the European Research Council (ERC) Horizon Europe under grant agreement number 101044048, from the Belgian Research Foundation Flanders (FWO) Odysseus program under grant number G0H9218N, from FWO grant G077822N, and from KU Leuven C1 grant BRAVE C16/23/009. The authors would also like to thank previous members of the KUL-EQUATION group for their earlier contributions, and the referee for their useful comments on the manuscript. Finally, JS extends his personal thanks to (the both retiring) Jo Puls and Stan Owocki, for teaching him lots of stuff about line-driven winds over the years. The following packages were used to analyze the data: {\fontfamily{qcr}\selectfont NumPy} \citep{harris_2020}, {\fontfamily{qcr}\selectfont SciPy} \citep{virtanen_2020}, {\fontfamily{qcr}\selectfont matplotlib} \citep{hunter_2007}.

\end{acknowledgements}

\bibliographystyle{aa}
\bibliography{references_ostar} 

\begin{appendix}

\section{NLTE effects}

We here evaluate potential NLTE effects on the line force multipliers computed by \href{https://lime.ster.kuleuven.be/}{LIME}. For this purpose we use the method applied in the line-driven wind statistics paper by \citet{puls_2000} for computing the ionisation balance and occupation numbers. This is an approximate NLTE method that modifies the Saha-Boltzmann relations by accounting explicitly for different radiation and gas temperatures, $T_{\rm rad}$ and $T_{\rm gas}$, as well as for a \textit{modified} spherical dilution factor $W$ of the illuminating radiation field (as defined in, e.g., Eq. 10 of \citealt{lucy_1993}). In comparison to the standard dilution factor, note that this modified $W$ contains an optical depth correction such that, by definition, $W=1$ at the stellar surface in order to obtain a smooth transition to LTE conditions. This approximate method (or variants of it) has been quite extensively and successfully applied before in a range of line-driven wind studies (for example by \citealt{Abbott82, lucy_1985, lucy_1993, puls_2000, petrenz_puls_2000}), and has also been calibrated against full NLTE calculations in \citet{puls_2005}. We here follow \citet{puls_2000} in using a frequency-independent radiation temperature and a Planckian radiation field, enabling straightforward and direct comparisons to results in the main paper. As discussed in \citet{puls_2000} (their Sect. 4.2.8), previous tests using more realistic photospheric fluxes generally show small effects, with the possible exception of cooler B-stars (see also \citealt{petrenz_puls_2000}). As we currently cut application of \href{https://lime.ster.kuleuven.be/}{LIME} at $T_{\rm eff} = 18$ kK we adhere to Planckian fluxes in this paper, deferring further studies of the influence from different irradiation fluxes to future work (which in any case would involve computations of newer and much larger grids of underlying model atmospheres in order to allow for general  applications).

\begin{figure*}
 \centering
 \includegraphics[width=18cm]{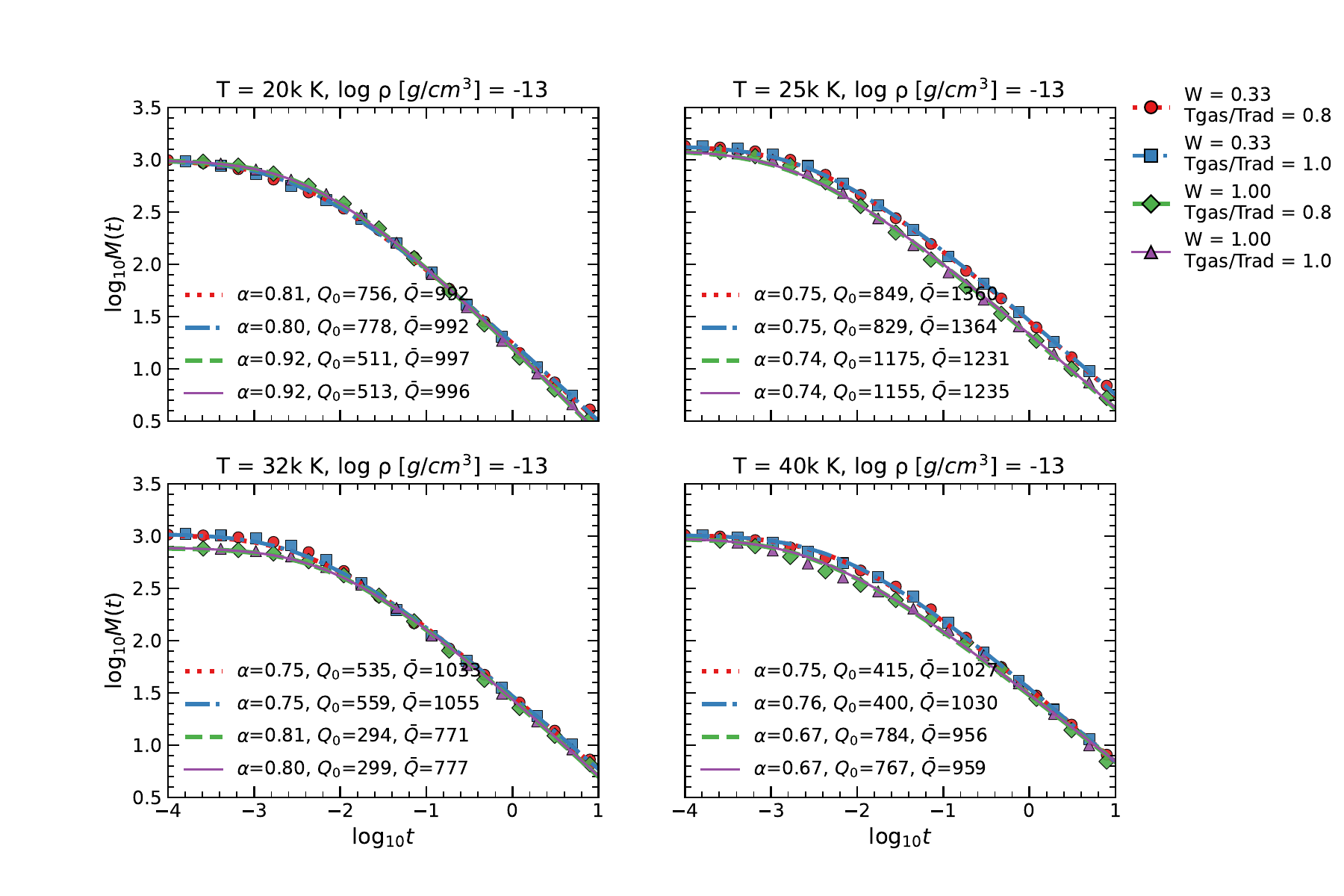}
      \caption{Force multipliers $M(t)$ computed for the densities and temperatures shown at top of each plot. Each plot shows four different $M(t)$ curves and corresponding fits of $\bar{Q}$, $Q_0$, and $\alpha$, computed for the different combinations of dilution factor $W$, radiation temperature $T_{\rm rad}$, and gas temperature $T_{\rm gas}$, displayed in the colour bar to the right. The temperatures $T$ at the top of each figure refer to $T_{\rm rad}$.}
  \label{Fig:cmp2}
\end{figure*}

Fig. \ref{Fig:cmp2} shows force multipliers $M(t)$ and fitted line-force parameters $\bar{Q}$, $Q_0$, and $\alpha$ computed with the new method. For a typical critical point density, results for four different gas temperatures are shown. $T_{\rm gas} = T_{\rm rad}$ and $W =1$ here correspond to the LTE approximation used in LIME, and we have verified that results are indeed identical. We test this against the ratio $T_{\rm gas}/T_{\rm rad} = 0.8$, also used by \citet{puls_2000}, and for $W =1$ as well as a very diluted radiation field $W = 1/3$ as in table 2 of \citealt{puls_2000}. From the figure we see that effects upon derived line force parameters are insignificant for all combinations with $W =1$ and remain generally small also for $W = 1/3$ (e.g., changes in $\alpha < 0.1$ throughout), albeit with some moderately large changes in $Q_0$ present. Most importantly though, corresponding effects upon predicted mass-loss rates remain modest. Note also that application of the low $W = 1/3$ means that the critical point would be shifted away from the stellar surface, which is not what our method here assumes and generally goes against the trend for line-driven wind models including the finite stellar disk (and even more so for models based on co-moving frame radiative transfer, see below).   

The relatively small NLTE effects found here are quite interesting in view of the general consensus that such effects are always very important in hot star atmospheres. But as discussed also in the main paper, this is particularly true further out in the wind where the radiation field is very diluted. On the other hand, near the star where the mass-loss rate is set, gas and radiation temperatures remain relatively similar and the radiation field is only modestly diluted. It is still true that a frequency-dependent radiation temperature and NLTE calculations are critical for certain population numbers (e.g., the HeII ground state, see \citealt{puls_2005}) and the formation of specific spectral lines, but the results presented here suggest that for the cumulative line acceleration such effects may perhaps play a somewhat smaller role than previously thought, at least near the stellar surface. Since the critical point is even closer to the surface (at the gas sonic point) in non-Sobolev stationary wind models based on co-moving frame line radiative transfer \citep{sander_2017, sundqvist_bjorklund_2019}, this may at least partly explain why the simple semi-analytic LTE method presented in this paper actually gives results that, in a general sense, agree rather well with these seemingly much more sophisticated simulations. Indeed, this was actually suggested already by the small set of full hydrodynamical LTE O-star line-driven wind simulations by \citet{luka_2022}, who found overall good agreement with the mass-loss rates predicted by complete NLTE models (their Fig. 7).

\end{appendix}

\end{document}